\documentclass[conference,10pt]{IEEEtran}
\usepackage{epsfig,rotating,setspace,latexsym,amsmath,epsf,amssymb,amsfonts,bm,theorem,subfigure,epstopdf}
\usepackage{cite,authblk}
\usepackage{bbm}
\usepackage{algorithmic,algorithm}

\IEEEoverridecommandlockouts

\allowdisplaybreaks
\usepackage{color}

\title{Sending Information Through Status Updates\thanks{This work was supported by NSF Grants CCF 14-22111/14-22347, CNS 15-26608/15-26165.}}

\author[1]{Abdulrahman Baknina}
\author[2]{Omur Ozel}
\author[3]{Jing Yang}
\author[1]{Sennur Ulukus}
\author[3]{Aylin Yener}
\affil[1]{\normalsize Department of Electrical and Computer Engineering, University of Maryland, College Park, MD 20742}
\affil[2]{\normalsize Department of Electrical and Computer Engineering, Carnegie Mellon University, Pittsburgh, PA 15213}
\affil[3]{\normalsize Department of Electrical Engineering, The Pennsylvania State University, University Park, PA 16802}

\begin{document}

\IEEEoverridecommandlockouts
\maketitle

\begin{abstract}
We consider an energy harvesting transmitter sending status updates regarding a physical phenomenon it observes to a receiver. Different from the existing literature, we consider a scenario where the status updates carry information about an independent message. The transmitter encodes this message into the timings of the status updates. The receiver needs to extract this encoded information, as well as update the status of the observed phenomenon. The timings of the status updates, therefore, determine both the age of information (AoI) and the message rate (rate). We study the tradeoff between the achievable message rate and the achievable average AoI. We propose several achievable schemes and compare their rate-AoI performances.
\end{abstract}

\section{Introduction}

We consider an energy harvesting transmitter sending status updates to a receiver via status update packets. Each status update packet requires a unit of energy; and the transmitter harvests energy stochastically over time, one unit at a time, at random times.\footnote{Energy requirements and energy harvests are normalized.} In order to minimize the age of information (AoI), the transmitter needs to send frequent and regular (over time) status updates, however, the frequency and regularity of the updates are constrained by the stochastic energy arrival process, which is known only causally at the transmitter.

\begin{figure}[t]
	\centerline{\includegraphics[width=.7\columnwidth]{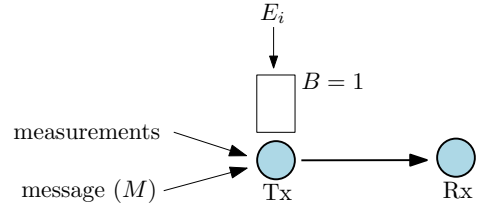}}
	\caption{An energy harvesting transmitter with a finite-sized battery, that sends status updates and independent information to a receiver.}
	\label{sys_model}
	\vspace{-0.5cm}
\end{figure}

In this paper, different from the existing literature, we consider the scenario where the timings of the status updates also carry an independent message; see Fig.~\ref{sys_model}. In order to obtain a tractable formulation, we consider an abstraction where the physical channel is noiseless and the transmitter has a battery of unit size. Intuitively, as will be clarified shortly, there is a tradeoff between the AoI and the rate of the message. Our goal in this paper is to characterize this tradeoff.

For this scenario, under causal (i.e., online) knowledge of energy arrivals, \cite{wu2017optimal_ieee} has determined that, in order to minimize the long-term average AoI, the transmitter needs to apply a \emph{threshold based} policy: There exists a fixed and deterministic threshold $\tau_0$ such that if an energy arrives sooner than $\tau_0$ seconds since the last update, the transmitter waits until $\tau_0$ and sends the update packet; on the other hand, if it has been more than $\tau_0$ seconds since the last update, the transmitter sends an update packet right away when an energy arrives.

On the other hand, again for this scenario, \cite{tutuncuoglu2017binary} has considered the information-theoretic capacity of this energy harvesting channel. The main information-theoretic challenge arises due to having a state-dependent channel (where the state is the energy availability), time-correlation introduced in the state due to the existence of a battery at the transmitter where energy can be saved and used later, and the unavailability of the state information at the receiver. Reference \cite{tutuncuoglu2017binary} converts the problem from regular channel uses to a timing channel and obtains the capacity in terms of some auxiliary random variables using a bits through queues approach as in \cite{anantharam1996bits}.

Sending information necessarily requires the transmitter to send out a packet after a random amount of time following an energy arrival in \cite{tutuncuoglu2017binary}, whereas minimizing AoI requires the transmitter to apply a deterministic threshold based policy in \cite{wu2017optimal_ieee}. Note that in \cite{wu2017optimal_ieee}, the transmitter sends a packet either at a deterministic time $\tau_0$ after an energy arrival, or right at the time of an energy arrival, thus, it cannot send any rate with the packet timings even though it minimizes the AoI. This is the main source of the tension between AoI minimization and information rate maximization; and is the subject of this paper.

In this paper, we first present a general tradeoff region between the achievable AoI and the achievable information rate. We then consider the class of renewal policies in which the system action depends only on the most recent transmission. Within this class of policies, we first propose policies that determine the next transmission instant as a function of the time difference between the most recent energy arrival and the most recent status update. We then consider simpler policies which we call {\it separable} policies. These policies separate the update decision and information transmission in an additive manner: When an energy arrives, the transmitter decides when to update, neglecting the information transmission; once the transmitter decides to send an update, it then encodes the message on top of that update timing. For all the policies, we derive the average achievable AoI and the achievable rate. We then compare the tradeoff regions of these policies. We observe numerically that the first class of policies achieve better tradeoff regions. We also observe that as the value of the average energy arrival increases, policies perform similarly.

{\it Related Work:} Minimizing the AoI has been studied in many different settings, including settings with no energy constraints \cite{kaul2012real, kaul2012status, yates2012real, kam2013age, costa2014age, sun2016update, kosta2017age, bedewy2017age, parag2017real, yates2017timely} and settings with energy constraints in \emph{offline} and \emph{online} energy harvesting models \cite{yates2015lazy, bacinoglu2015age, wu2017optimal_ieee, arafa2017age2, arafa2017age}. Energy harvesting communication systems have been extensively studied in scheduling-theoretic and information-theoretic settings, for example, \emph{offline scheduling} in single-user and multi-user settings have been considered in \cite{jingP2P, kayaEmax, omurFade, ruiZhangEH, jingBC, jingMAC, aggarwalPmax, kaya-interference, gunduzLoss}, \emph{online scheduling} has been considered in \cite{omurFade, wang2013simplicity, khuzani2014online, shaviv2015universally, onlineBC-journal,  wiopt_online_MAC, onlineMAC}, and information-theoretic limits have been considered in \cite{ozel2012achieving, mao2017capacity, shaviv2016capacity, jog2016geometric, tutuncuoglu2017binary}.

\section{System Model}\label{sec_sys_model}
We consider a noiseless binary energy harvesting channel where the transmitter sends status updates and an independent message simultaneously as in Fig.~\ref{sys_model}. The transmitter has a unit size battery, i.e., $B=1$. Energy arrivals are known causally at the transmitter and are distributed according to an i.i.d.~Bernoulli distribution with parameter $q$, i.e., $\mathbb{P}[E_i=1]=1-\mathbb{P}[E_i=0]=q$. Hence, the inter-arrival times between the energy arrivals, denoted as $\tau_i \in\{1,2,\cdots\}$, are geometric with parameter $q$. Each transmission costs unit energy; thus, when the transmitter sends an update, its battery is depleted. The timings of the transmitted updates determine the average AoI and the message rate.

\begin{figure}[t]
	\centerline{\includegraphics[width=1.0\columnwidth]{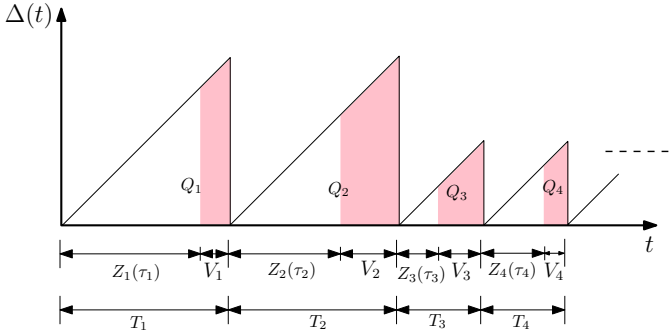}}
	\caption{An example evolution of instantaneous AoI.}
	\label{age_info_diagram}
	\vspace{-0.5cm}
\end{figure}

The instantaneous AoI is given by
\begin{align}
\Delta(t)=t-u(t)
\end{align}
where $u(t)$ is the time stamp of the latest received status update packet and $t$ is the current time.
An example evolution of the AoI is shown in Fig.~\ref{age_info_diagram}. The average long-term AoI is
\begin{align}
\Delta =& \limsup_{n\rightarrow \infty} \mathbb{E}\left[\frac{\sum_{j=1}^{n}Q_j}{\sum_{j=1}^{n}T_j}\right] \\
=&\limsup_{n\rightarrow \infty}  \mathbb{E}\left[ \frac{\sum_{j=1}^{n}T_j^2}{2\sum_{j=1}^{n}T_j}\right]
\end{align}
where $T_i$ is the duration between two updates, $Q_j=T_j^2/2$ is the total accumulated age between two updates represented by the area (see Fig.~\ref{age_info_diagram}), and the expectation is over the energy arrivals and possible randomness in the transmission decisions. Then, the minimum AoI is given by
\begin{align}
\Delta^*= \inf_{\pi \in \Pi}\Delta
=\inf_{\pi \in \Pi} \limsup_{n\rightarrow \infty}  \mathbb{E}\left[ \frac{\sum_{j=1}^{n}T_j^2}{2\sum_{j=1}^{n}T_j} \right]
\end{align}
where $\Pi$ is the set of all feasible policies. Since the transmitter is equipped with a unit-sized battery and due to energy causality \cite{jingP2P}, we have $T_i \geq \tau_i$. Note that due to the memoryless property of the geometric distribution, we assume without loss of generality, that $\tau_i$ is the time from the instant of the previous update and not the time from the instant of the previous energy arrival.

\begin{figure}[t]
	\centerline{\includegraphics[width=1.0\columnwidth]{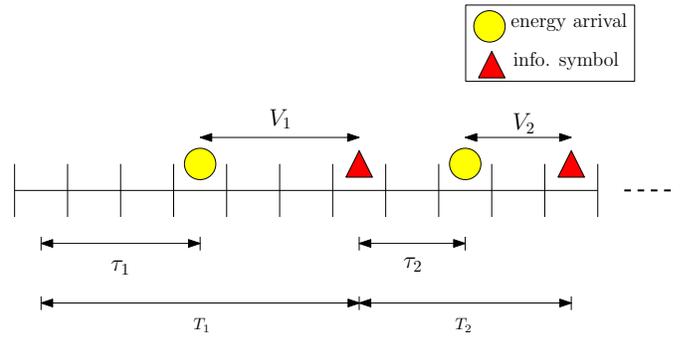}}
	\caption{Sending information through a timing channel.}
	\label{sending_inf_thr_timing}
	\vspace{-0.5cm}
\end{figure}

To send information through the timings of the status updates, we consider the model studied in \cite[Section~V.A]{tutuncuoglu2017binary}. Thus, here, we assume the knowledge of the energy arrival instants causally at the transmitter and the receiver. The information in the time duration $T_i$ is carried by the random variable $V_i \in\{0,1,\cdots\}$ where we have here $T_i=\tau_i+V_i$, see Fig.~\ref{sending_inf_thr_timing}. The achievable rate of this timing channel is \cite{tutuncuoglu2017binary},
\begin{align}
R& =  \liminf_n \sup_{p(V^n|\tau^n)} \frac{I(T^n;V^n|\tau^n)}{\sum_{i=1}^{n} \mathbb{E}[V_i]+\mathbb{E}[\tau_i]} \\
&=  \liminf_n \sup_{p(V^n|\tau^n)} \frac{H(V^n|\tau^n)}{\sum_{i=1}^{n} \mathbb{E}[V_i]+\mathbb{E}[\tau_i]} \label{eq_max_rate}
\end{align}
where the second equality follows since $H(V^n|\tau^n,T^n)=0$.

We denote the AoI-rate tradeoff region by the tuple $(\text{AoI}(r),r)$, where $r$ is the achievable rate and $\text{AoI}(r)$ is the minimum achievable AoI given that a message rate of at least $r$ is achievable,
\begin{align}
\text{AoI}(r)= \inf_{\mathcal{M}} \limsup_{n\rightarrow \infty}  \mathbb{E}\left[\frac{\sum_{j=1}^{n}T_j^2}{2\sum_{j=1}^{n}T_j} \right]
\end{align}
where $\mathcal{M}$ is defined as
\begin{align}
\!\! \mathcal{M}\!\!= \!\!\left\{ \!\! \{T_i\}_{i=1}^{\infty} \Bigg|  T_i \geq \tau_i, \liminf_n \!\!\!\! \sup_{p(V^n|\tau^n)} \frac{H(V^n|\tau^n)}{\sum_{i=1}^{n} \mathbb{E}[V_i]+\mathbb{E}[\tau_i]} \!\geq r  \!\right\}\!
\end{align}
where $V^n$ denotes $(V_1,\cdots,V_n)$ and similarly for $\tau^n$. An alternate characterization for the tradeoff region can also be done using the tuple $(\alpha, R(\alpha))$ where the achievable AoI is equal to $\alpha$ and $R(\alpha)$ is the maximum achievable information rate given that the AoI is no more than $\alpha$.

\section{Achievable Tradeoff Regions}\label{sec_sim_status_upd_and_info}
In this section, we consider several achievable schemes. All considered achievable schemes belong to the class of renewal policies. A renewal policy is a policy in which the action $T_i$ at time $i$ is a function of only the current energy arrival instant $\tau_i$. The long-term average AoI under renewal policies is,
\begin{align}\label{aoi-renewal}
\Delta = \limsup_{n\rightarrow \infty}  \mathbb{E}\left[\frac{\sum_{j=1}^{n}T_j^2}{2\sum_{j=1}^{n}T_j} \right] 
= \frac{\mathbb{E}[T_i^2] }{2\mathbb{E}[T_i]}
\end{align}
which results from renewal reward theory \cite[Theorem~3.6.1]{ross1996stochastic}. Since we use renewal policies and $\tau_i$ is i.i.d., hereafter, we drop the subscript $i$ in the random variables. Then, the maximum achievable information rate in (\ref{eq_max_rate}) reduces to,
\begin{align}
R=\max_{p(v|\tau)} \frac{H(V|\tau)}{\mathbb{E}[V]+\mathbb{E}[\tau]}
\end{align}
and the AoI in (\ref{aoi-renewal}) reduces to 
\begin{align}
\Delta =  \frac{\mathbb{E}[T^2] }{2\mathbb{E}[T]} = \frac{\mathbb{E}[(V+\tau)^2]}{2 \mathbb{E}[V+\tau]}
\end{align}

Next, we present our achievable schemes. In the first scheme, information transmission is adapted to the timing of energy arrivals: If it takes a long time for energy to arrive, the transmitter tends to transmit less information and if energy arrives early, the transmitter tends to transmit more information. This scheme fully adapts to the timings of the energy arrivals, but this comes at the cost of high computational complexity. We then relax the adaptation into just two regions, divided by a threshold $c$: If energy arrives in less than $c$ slots, we transmit the information using a geometric distribution with parameter $p_b$, and if energy arrives in more than $c$ slots, we transmit the information using another geometric random variable with parameter $p_a$. The choice of a geometric random variable for $V$ here and hereafter is motivated by the fact that it maximizes the information rate when the energy arrival timings are known at the receiver;  see \cite[Section~V.A]{tutuncuoglu2017binary}.

In the previous schemes, the instantaneous information rate depends on the timings of energy arrivals. We next relax this assumption and assume that the instantaneous information rate is fixed and independent of timings of energy arrivals. We call such policies \emph{separable} policies. In these policies, the transmitter has two separate decision blocks: The first block is for the status update which takes the decision depending on the timing of the energy arrival, and the second block is for encoding the desired message on top of the timings of these updates. This is similar in spirit to super-position coding. In the first separable policy, the update decision is a threshold based function inspired by \cite{wu2017optimal_ieee}: if the energy arrives before a threshold $\tau_0$, the update block decides to update at $\tau_0$ and if the energy arrives after $\tau_0$, the update block decides to update immediately. The information block does not generate the update immediately, but adds a geometric random variable to carry the information in the timing on top of the timing decided by the update block. In the second separable policy, which we call zero-wait policy, the update block decides to update in the channel use immediately after an energy arrival.

\subsection{Energy Timing Adaptive Transmission Policy (ETATP)}
In this policy, the information which is carried in $V$ is a (random) function of the energy arrival realization $\tau$. This is the most general case under renewal policies. The optimal tradeoff can be obtained by solving the following problem
\begin{align}
\min_{p(v|\tau)} \quad & \frac{\mathbb{E}[(V+\tau)^2]}{2 \mathbb{E}[V+\tau]} \nonumber \\
\mbox{s.t.} \quad & \frac{H(V|\tau)}{\mathbb{E}[V]+\mathbb{E}[\tau]} \geq r
\end{align}
The maximum possible value for $r$ is equal to $r^*= \max_{p(v|\tau)}\frac{H(V|\tau)}{\mathbb{E}[V]+\mathbb{E}[\tau]}$. The solution of this problem can be found by considering the following alternative problem which gives the same tradeoff region
\begin{align}
\label{eq_best_ach}
\max_{p(v|\tau),m} \quad & \frac{H(V|\tau)}{m} \nonumber\\
\mbox{s.t.} \quad & \mathbb{E}[(V+\tau)^2] \leq 2\alpha\mathbb{E}[V+\tau] \nonumber \\
& \mathbb{E}[V+\tau] = m
\end{align}
For a fixed $m$, problem (\ref{eq_best_ach}) is concave in $p(v|\tau)$ and can be solved efficiently. Then, to obtain the entire tradeoff region, we sweep over all possible values of the parameter $\alpha$ (which are all possible values of the AoI). The solution for (\ref{eq_best_ach}) is found numerically by optimizing over all possible conditional pmfs $p(v|\tau)$ for each value of $m$. Then, we use line search to search for the optimal $m$. All this, has to be repeated for all possible values of the AoI $\alpha$. Finding the optimal solution for (\ref{eq_best_ach}) has a high complexity, hence, we propose the following policy which reduces this complexity significantly, and at the same time adapts to the timing of the energy arrivals to the extent possible within this set of policies.

\subsection{Simplified ETATP}
In this policy, we simplify the form of the dependence of the transmission on the timings of energy arrivals significantly. The transmitter waits until an energy arrives, if the energy takes more than $c$ slots since the last update, we transmit the information using a geometric random variable with probability of success $p_b$, otherwise the transmitter transmits the information using a geometric random variable with probability of success $p_a$, i.e., the transmitter chooses $p(v|\tau)$ as follows
\begin{align}
p(v|\tau)=
\begin{cases}
p_b (1-p_b)^{v-1},\ \  \tau<c\\
p_a (1-p_a)^{v-1},\ \  \tau\geq c
\end{cases}
, \ v=1,2,\cdots
\end{align}
In this case, $p_a$, $p_b$ and $c$ are the variables over which the optimization is performed.
The average achieved information rate as a function of $p_a$, $p_b$ and $c$ can be obtained as,
\begin{align}
R=\frac{\frac{H_2(p_b)}{p_b}(1-(1-q)^c)+\frac{H_2(p_a)}{p_a}(1-q)^c}{\mathbb{E}[\tau] +
	\mathbb{E}[V]}
\end{align}
where $\mathbb{E}[V]$ is equal to
\begin{align}
\mathbb{E}[V]= \frac{(1-p_b)}{p_b}(1-(1-q)^c) +\frac{(1-p_a)}{p_a}(1-q)^c
\end{align}
Now, we can calculate the average AoI with this policy as,
\begin{align}
\Delta = \frac{\mathbb{E}[(\tau+V)^2] }{2\mathbb{E}[\tau+V]}  =\frac{ \frac{2-q}{q^2}+\mathbb{E}[V^2] +2\mathbb{E}[\tau V] }{2\mathbb{E}[\tau]+2\mathbb{E}[V]}
\end{align}
where we have $\mathbb{E}[V^2]$ as
\begin{align}
\mathbb{E}[V^2]=&\left(\frac{2+p_b^2-3p_b}{p_b^2}\right) \left(1-(1-q)^c\right) \nonumber \\
&+ \left(\frac{2+p_a^2-3p_a}{p_a^2}\right) (1-q)^c
\end{align}
and $\mathbb{E}[\tau V]$ as
\begin{align}
\mathbb{E}[\tau V]=&  \frac{(1-p_b)}{p_b}\left(\frac{1}{q}(1-(1-q)^{c+1}) - (c+1)(1-q)^c\right) \nonumber \\
& +\frac{(1-p_a)}{p_a}\left(\frac{(1-q)^{c+1}}{q}+c(1-q)^{c-1}\right)
\end{align}

This schemes is simpler than the general class of ETATP; still, we need to search for the optimal $p_a$, $p_b$ and $c$. We reduce this complexity further in the next policy.

\subsection{Threshold Based Transmission Policy}
We now present the first separable policy. In this policy, we assume that $T=Z(\tau)+V$, where the information is still carried only in $V$; see Fig.~\ref{age_info_diagram}. $Z(\tau)$ is the duration the transmitter decides to wait in order to minimize the AoI, while $V$ is the duration the transmitter decides to wait to add information in the timing of the update. $Z(\tau)$ and $V$ are independent which implies that $H(V|Z(\tau))=H(V|\tau)=H(V)$. The duration $Z(\tau)$ is determined according to a threshold policy as follows,
\begin{align}
Z(\tau)= \tau U(\tau -\tau_0 )+\tau_0  U(\tau_0 -\tau -1)
\end{align}

The optimal value of $\tau_0 $ is yet to be determined and is an optimization variable. The optimal value of $\tau_0$ is to be calculated and, thus, known both at the transmitter and the receiver; hence, this threshold policy is a deterministic policy. This ensures that we still have $H(V^n|\tau^n,T^n)=0$, which is consistent with (\ref{eq_max_rate}). We then choose $V$ to be a geometric random variable with parameter $p$. The tradeoff region can then be written as,
\begin{align}\label{prob_original}
\min_{T(\tau),p} \quad & \frac{\mathbb{E}[(Z(\tau)+V)^2] }{2\mathbb{E}[Z(\tau)+V]}     \nonumber \\
\mbox{s.t.} \quad &  Z(\tau) \geq \tau \nonumber \\
&r \leq \frac{H_2(p)/p}{(1-p)/p+\mathbb{E}[Z(\tau)]}
\end{align}
where $r$ is a fixed positive number. The feasible values of $r$ are in $[0,r*]$ where $r^*$ is equal to $r^*= \max_{p \in [0,1] } \frac{H_2(p)/p}{(1-p)/p+\mathbb{E}[\tau]}$. This follows because the smallest value that $Z(\tau)$ can take is equal to $\tau$. The optimization problem in this case becomes a function of only $\tau_0 $ and $p$.

We now need to calculate $\mathbb{E}[Z(\tau  )]$ and $\mathbb{E}[Z^2(\tau  )]$. We calculate $\mathbb{E}[Z(\tau  )]$ as follows,
\begin{align}
\mathbb{E}[Z(\tau  )]=&  (1-q)^{\tau_0  } +  \frac{(1-q)^{\tau_0 +1}}{q} +\tau_0
\end{align}
and we calculate $\mathbb{E}[Z^2(\tau )]$ as follows,
\begin{align}
\mathbb{E}[Z^2(\tau  )]=&  \left(\frac{2-3q}{q^2} \right) (1\!-\!q)^{\tau_0  } + 2(\tau_0 \!+\!1)(1\!-\!q)^{\tau_0  } \nonumber \\
&+2(\tau_0 +1) \frac{(1-q)^{\tau_0 +1}}{q}+\tau_0 ^2
\end{align}
Finally, we note that in this case $\mathbb{E}[V^2]$ is equal to,
\begin{align}
\mathbb{E}[V^2]= \frac{2+p^2-3p}{p^2}
\end{align}
Substituting these quantities in the above optimization problem and solving for $p$ and $\tau_0 $ jointly gives the solution.

\subsection{Zero-Wait Transmission Policy}
This policy is similar to the threshold based policy, with one difference: The update block does not wait after an energy arrives, instead, it decides to update right away, i.e., $Z(\tau)=\tau$. Hence, the tradeoff region can be obtained by solving,
\begin{align}
\min_{p} \quad & \frac{\mathbb{E}[(\tau+V)^2] }{2\mathbb{E}[\tau+V]}     \nonumber \\
\mbox{s.t.} \quad & r \leq \frac{H_2(p)/p}{(1-p)/p+\mathbb{E}[\tau]}
\end{align}
We can then calculate $\mathbb{E}[(\tau+V)^2] =\mathbb{E}[\tau^2+V^2+2V\tau] $, where $V$ and $\tau$ are independent as the message is independent of the energy arrivals. Since $\tau$ is geometric $\mathbb{E}[\tau^2]=\frac{2-q}{q^2}$.
This optimization problem is a function of only a single variable $p$. This problem is solved by line search over $p\in[0,1]$.

\section{Numerical Results}\label{sec_numerical_results}
\begin{figure}[t]
	\centerline{\includegraphics[width=0.9\columnwidth]{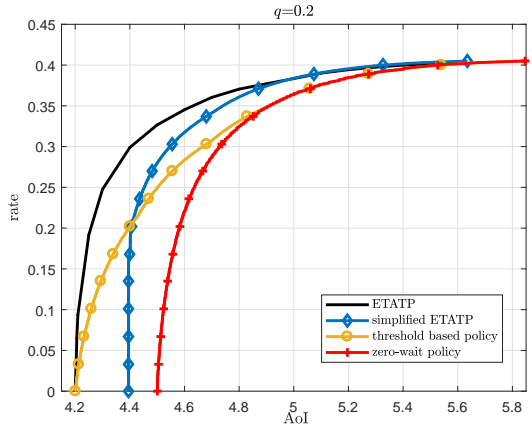}}
    \vspace{-0.2cm}
	\caption{The rate-AoI tradeoff region for $q=0.2$.}
	\label{res1}
	\vspace{-0.4cm}
\end{figure}

\begin{figure}[t]
	\centerline{\includegraphics[width=0.9\columnwidth]{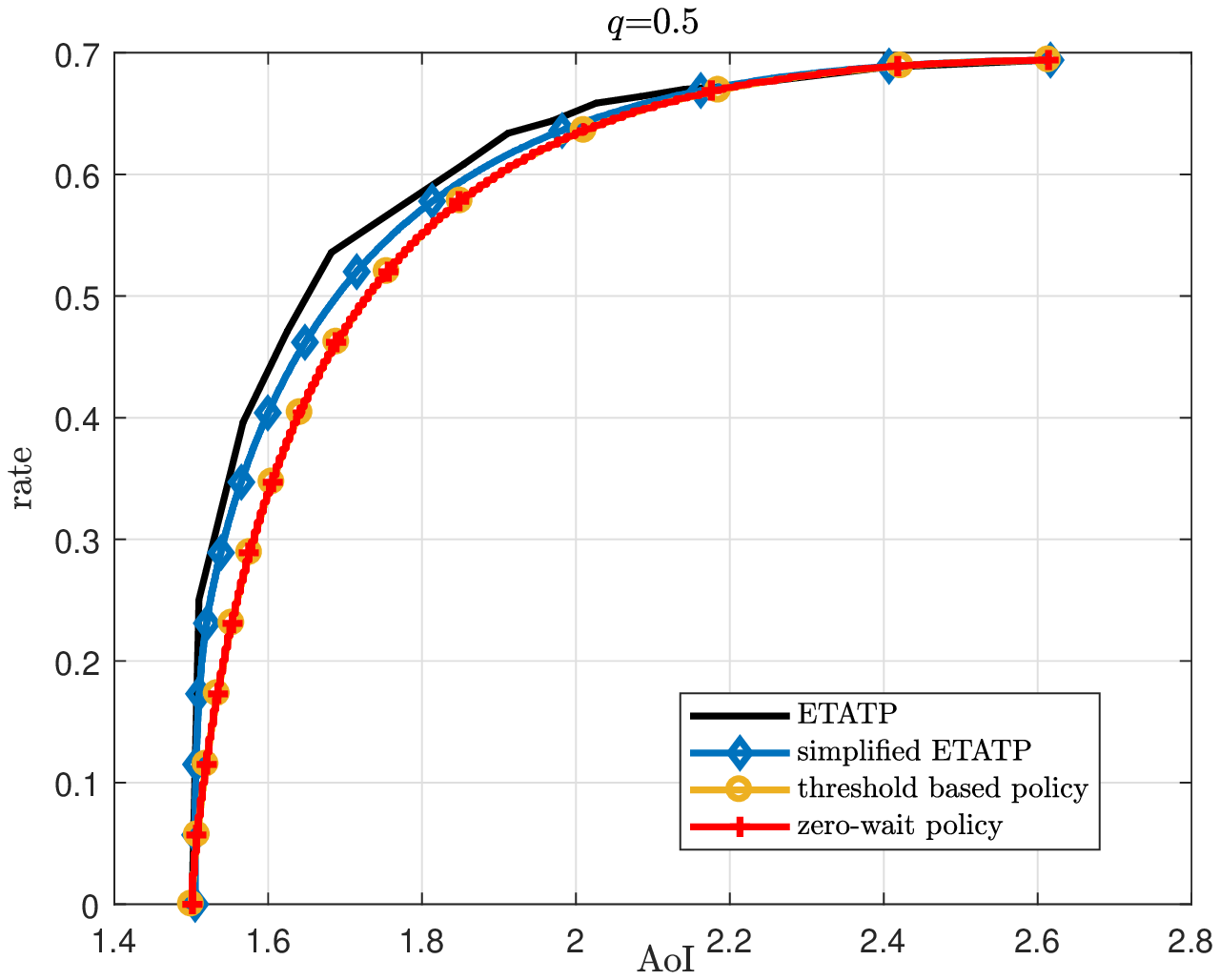}}
    \vspace{-0.2cm}
	\caption{The rate-AoI tradeoff region for $q=0.5$.}
	\label{res2}
	\vspace{-0.7cm}
\end{figure}

\begin{figure}[t]
	\centerline{\includegraphics[width=0.9\columnwidth]{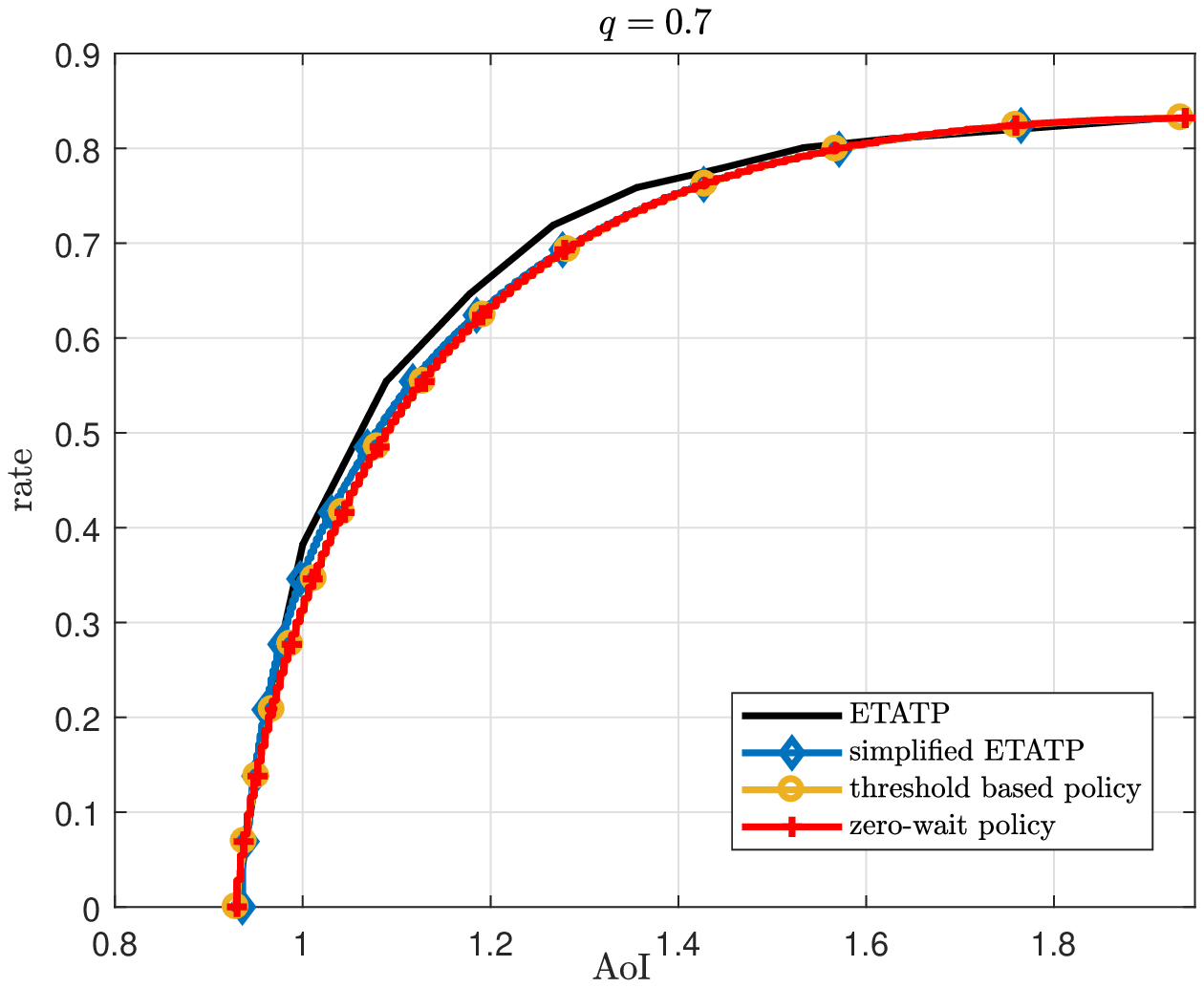}}
    \vspace{-0.2cm}
	\caption{The rate-AoI tradeoff region for $q=0.7$.}
	\label{res3}
	\vspace{-0.6cm}
\end{figure}

Here, we compare the tradeoff regions resulting from the proposed schemes. We plot these regions in Figs.~\ref{res1}-\ref{res3} for different values of average energy arrivals, namely, $q=0.2$, $q=0.5$ and $q=0.7$. For low values of $q$, as for $q=0.2$ in Fig.~\ref{res1}, there is a significant gap between the performance of ETATP and the simplified schemes. For this value of $q$, in most of the region, simplified ETATP performs better than the threshold and zero-wait policies. As the value of $q$ increases as shown in Fig.~\ref{res2} and Fig.~\ref{res3}, the gap between the performance of the different policies decreases significantly. In Fig.~\ref{res2}, the threshold and zero-wait policies overlap. In Fig.~\ref{res3}, simplified ETATP, threshold and zero-wait policies overlap. In all cases, zero-wait policy performs the worst. This is consistent with early results e.g., \cite{kaul2012real}, early results in the context of energy harvesting e.g., \cite{yates2015lazy, bacinoglu2015age}, and recent results \cite{wu2017optimal_ieee, arafa2017age2, arafa2017age}, where updating as soon as one can is not optimum.

\bibliographystyle{unsrt}
\bibliography{IEEEabrv,myLibrary}

\end{document}